\documentclass[a4paper]{article}
\usepackage{amsmath,cite}
\begin{document}
\begin{titlepage}

\begin{flushright}
BUTP/2000-6\\
\end{flushright}
\vspace{1cm}

\begin{center}
{\LARGE The construction of generalized Dirac operators on the lattice}
\footnote{Work supported in part by Schweizerischer Nationalfonds}
 
\vspace{1cm}
{\large
Peter Hasenfratz, 
Simon Hauswirth,
Kieran Holland, \\
Thomas J\"org, 
Ferenc Niedermayer,
Urs Wenger
}
\\
\vspace{0.5cm}
Institute for Theoretical Physics \\
University of Bern \\
Sidlerstrasse 5, CH-3012 Bern, Switzerland

\vspace{0.5cm}
March, 2000 \\ \vspace*{0.5cm}

{\bf Abstract}
\end{center}
\vspace{-5mm}

\begin{quote}
We discuss the steps to construct Dirac operators which have arbitrary fermion
offsets, gauge paths, a general structure in Dirac space and satisfy the basic
symmetries (gauge symmetry, hermiticity condition, charge conjugation,
hypercubic rotations and reflections) on the lattice. We give an extensive set
of examples and offer help to add further structures.
\end{quote}
 
\vfill
\end{titlepage}

\section{Introduction}

Dirac operators which satisfy the Ginsparg-Wilson (GW) relation
\begin{equation}
\label{gw}
\gamma^5 D^{-1} + D^{-1} \gamma^5 = \gamma^5 2 R\,,
\end{equation}
where $R$ is a local operator trivial in Dirac space,
define lattice regularized theories whose chiral properties are the same as
those of the corresponding formal continuum theories \cite{GW,Be,Neu,ML,GEN}.
The relation in eq.~(\ref{gw}) implies an exact chiral symmetry 
on the lattice \cite{ML}. Together with the earlier results on domain 
wall \cite{DS} and overlap \cite{OVER} fermions, the recent developments 
provide a rather satisfactory understanding of global chiral symmetry 
on the lattice.
There is significant progress on chiral gauge theories also \cite{CHI}.

Two types of solutions to the GW relation are known: the fixed-point (FP)
Dirac operators of different renormalization group transformations \cite{Be}
and the Dirac operators obtained by using Neuberger's construction
\cite{Neu}. In the latter case one might even start with the simplest
nearest-neighbour Wilson operator. It is probable, however that starting with
a Dirac operator which satisfies the GW relation approximately will reduce the
numerical problems with the square root in Neuberger's construction and will
more than pay back the additional overhead\footnote{Starting with the Wilson
operator the end result after Neuberger's construction is local but very
broad  and has strong cut-off effects \cite{FP}. This gives additional
(and related) reasons to start with an operator with better properties.}. In
this case and, of course, in any parametrization of the FP operator one 
faces the problem of constructing a Dirac matrix with different fermion 
offsets, gauge paths and a sufficiently rich structure in Dirac space 
in such a way that the basic symmetries 
(gauge symmetry, $\gamma_5$ hermiticity, charge conjugation, 
hypercubic rotations and reflections) are satisfied.

This problem was addressed in $d=2$ in a recent paper by Gattringer and Hip
\cite{GH}. Considering gauge paths with length up to 2 and all 4 matrices
of the $d=2$ Clifford algebra they discussed and solved the symmetry
conditions. The parameters were fixed then by optimizing for the GW
relation. This program has been extended to $d=4$ in a follow-up
paper by Gattringer \cite{Ga}.
The first step, i.e. writing down an Ansatz for a generalized Dirac
operator which satisfies the basic symmetries, is however independent of the
method used to fix the parameters.

In this paper we discuss this kinematic problem in $d=4$. For any fermion
offset and for any of the 16 elements of the Clifford algebra we describe the
steps to find combinations of gauge paths which satisfy all the basic
symmetries. We construct explicitly paths for all the elements of the
Clifford algebra in the offsets of the hypercube. We show also how to
factorize the sum of paths (i.e. writing it as a product of sums) in such a
way that the computational problem is manageable even if the number of paths
is large.

We attempted to organize this paper in a way which makes it possible to use
the results without the need of repeating the straightforward but lengthy
algebra involved. Care was taken to introduce notations and to fix the
conventions. We explain with examples how to use the tables and the listed 
combinations of gauge field products in order to build up the matrix elements
of the Dirac operator.

In case the reader wants to consider offsets and paths
which do not enter our list, we offer an easy-to-use Maple code 
{\tt dirac.maple} which is included with this paper 
in the {\tt hep-lat} archive.
The user should specify a Clifford algebra element and a
gauge path, and the output will help to construct the corresponding
contribution to the Dirac matrix.

Let us add a few remarks at this point:\\
1) There are many ways to fix the parameters of the Ansatz. As opposed to
production runs this problem should be treated only once. It is useful to
invest effort here, since the choice will influence strongly the quality of
the results in simulations. We plan to optimize the parameters to the FP
action. \\
2) There are compelling reasons to use the elements of the Clifford algebra
beyond $1$ and $\gamma^\mu$ in the Dirac operator. For an
operator D satisfying the GW relation in eq.~(\ref{gw}), 
${\rm Tr}(\gamma^5 R D)$ is
the topological charge density \cite{Be,ML}, 
i.e the $\propto \gamma^5$ part of $D$ is
obviously important. Similarly, a $\propto \sigma^{\mu \nu}$ term is already
required by the leading Symanzik condition. \\
3) If the parametrization is good, the GW relation is approximately satisfied,
the operator under the square root in  Neuberger's construction is close to
1. An expansion is expected to converge very fast in this case.\\
4) The basic numerical operation in production runs
is $D\eta$, where $D$ is the parameterized Dirac matrix and $\eta$ 
is a vector. The matrix elements of $D$ should be precalculated before
the iteration starts. Using all the Clifford algebra elements and arbitrary 
gauge paths, the computational cost per offset of the operation 
$D\eta$ is a factor of $\sim 4$ higher than that of the Wilson action. 
Keeping the points of the hypercube without the outmost corners 
(65 offsets), say, the cost per iteration is increased by a factor 
of $4\times 65/8 \sim 30$ relative to Wilson Dirac operator.

\section{Symmetries of the Dirac operator}

We define the basis of the Clifford algebra as $\Gamma=1,\,\gamma_\mu,
\,i\sigma_{\mu  \nu},\,\gamma_5,\, \gamma_\mu \gamma_5$. 
We use the notation S,V,T,P and A for the scalar, vector, tensor, 
pseudoscalar and axial-vector elements of the Clifford algebra,
respectively. Notice that the tensor (T) and axial-vector (A) basis 
elements of the Clifford algebra are {\it anti-hermitian}.
It will later be useful to list the basis elements of the Clifford 
algebra by a single index, as $\Gamma_A$, $A=1,\ldots,16$. 
We choose the ordering 
\begin{equation}
\label{cord}
1,\gamma_1,\gamma_2,\gamma_3,\gamma_4, i\sigma_{12},i\sigma_{13},
i\sigma_{14},i\sigma_{23},i\sigma_{24},i\sigma_{34},\gamma_5,
\gamma_1\gamma_5,\gamma_2\gamma_5,\gamma_3\gamma_5,\gamma_4\gamma_5.
\end{equation}

Under charge conjugation, the Dirac matrices transform as
\begin{equation}
\label{cdef}
{\cal C} \gamma_\mu^T {\cal C}^{-1} = - \gamma_\mu \,.
\end{equation}

Having chosen the T and A basis elements of the Clifford algebra to be
anti-hermitian, all the basis elements have the property that 
\begin{equation}
\label{pleasant}
{\cal C}\gamma_5 \Gamma^* \gamma_5 {\cal C}^{-1}= \Gamma \,.
\end{equation}

We define a sign $\epsilon_{\Gamma}$ by the relation
\begin{equation}
\label{epsi}
\gamma_5 \Gamma^\dagger \gamma_5 =\epsilon_\Gamma \Gamma,
\end{equation}
which gives $\epsilon_{\rm S}=\epsilon_{\rm P}=\epsilon_{\rm A}=1$ and 
$\epsilon_{\rm V}=\epsilon_{\rm T}=-1$.

Under reflection of the coordinate axis $\eta$, the basis elements 
of the Clifford algebra are transformed as $\Gamma'={\cal P}_\eta \Gamma 
{\cal P}_\eta^{-1}$, where ${\cal P}_\eta = \gamma_{\eta}\gamma_5$.
(When $\Gamma$ is written in terms of Dirac matrices $\gamma_\mu$
this amounts simply to reversing the sign of $\gamma_\eta$.)
The group of reflections has $2^4=16$ elements.
Under a permutation $(1,2,3,4)\to (p_1,p_2,p_3,p_4)$ of
the coordinate axes the basis elements of the Clifford algebra
transform by replacing $\mu\to p_\mu$ and accordingly
$\gamma_5 \to \epsilon_{p_1 p_2 p_3 p_4} \gamma_5$.

The matrix elements of the Dirac operator $D$ we denote as
\begin{equation}
D(n,n';U)_{\alpha \alpha'}^{aa'}
\end{equation}
where $n$, $a$ and $\alpha$ refer to the coordinate, colour and Dirac
indices, respectively.
From now on, we suppress the colour and Dirac indices.
The gauge configuration $U$ appearing in $D$ is not necessarily
the original one entering the gauge action -- it could be a
smeared configuration.
Provided the smearing has appropriate symmetry properties
(as all conventional smearing schemes do) the constructions
in this paper remain valid, without any extra modifications.

The Dirac operator should satisfy the following symmetry requirements:

\noindent
{\bf gauge symmetry} \\
\noindent
Under the gauge transformation $U_\mu(n) \rightarrow  U_\mu(n)^g=
g(n)\,U_\mu(n)\,g(n+\hat{\mu})^\dagger$, where $g(n) \in {\rm SU}(N)$ 
we have
\begin{equation}
D(n,n';U) \rightarrow  \,D(n,n';U^g) =
g(n)\, D(n,n';U)  \,g(n')^\dagger .
\label{gau}
\end{equation}

\noindent
{\bf translation invariance} \\
\noindent
Translation symmetry requires that $D(n,n+r)$
depends on $n$ only through the $n$-dependence of the gauge fields.
There is no explicit $n$-dependence beyond that. In
particular, the coefficients in front of the different paths which enter
$D$ do not depend on $n$.

\noindent
{\bf hermiticity}
\begin{equation}
D(n,n';U) = \gamma^5 \, D(n',n,U)^\dagger \,\gamma^5,
\label{s1}
\end{equation}
where $\dagger$ is hermitian conjugation in colour and Dirac space.

\noindent
{\bf charge conjugation}
\begin{equation}
D(n,n';U) = {\cal C} \, D(n',n;U^*)^T \,{\cal C}^{-1} \,,
\label{s2}
\end{equation}
where $T$ is the transpose operation in colour and Dirac space.

It will be useful to combine eqs.~(\ref{s1},\ref{s2}) to obtain
\begin{equation}
\label{ss2}
D(n,n';U) = {\cal C} \gamma^5 \, D(n,n';U^*)^* \,\gamma^5 {\cal C}^{-1} \,.
\end{equation}

\noindent
{\bf reflection of the coordinate axis $\eta$} \\
\begin{equation}
D(n,n';U) = {\cal P}^{-1}_\eta \, D(\tilde{n},\tilde{n}';U^{{\cal P}_\eta})
\,{\cal P}_\eta \,,
\label{s3}
\end{equation}
where ${\cal P}_\eta = \gamma^{\eta}\gamma^5$ and $\tilde{n}_\nu=n_\nu$ if
$\nu \neq \eta$, while $\tilde{n}_\eta=-n_\eta$.
The reflected gauge field $U^{{\cal P}_\eta}$ is defined as
\begin{eqnarray}
\label{12}
U^{{\cal P}_\eta}_\eta(m) & = & U_\eta(\tilde{m}-\hat{\eta})^\dagger \,, \\
U^{{\cal P}_\eta}_\nu(m) & = & 
U_\nu(\tilde{m})\,,\qquad  \nu \neq \eta .\nonumber
\end{eqnarray} 

\noindent
{\bf permutation of the coordinate axes} \\
\noindent
These are defined in a straightforward way, by permuting the
Lorentz indices appearing in $D$.
Note that rotations by  $90^\circ$ on a hypercubic lattice can be replaced
by reflections and permutations of the coordinate axes.

\section{Constructing a matrix $D$ which satisfies the symmetries}

To describe a general Dirac operator in compact notations, 
it is convenient to introduce
the operator $\hat{U}_\mu$ of the parallel transport for direction $\mu$
\begin{equation}
\label{Uhat}
\left(\hat{U}_\mu\right)_{nn'}=U_\mu(n)\delta_{n+\hat{\mu},n'} \,,
\end{equation}
and analogously for the opposite direction
\begin{equation}
\label{Uhatd}
\left(\hat{U}_{-\mu}\right)_{nn'}=U_\mu(n-\hat{\mu})^\dagger
  \delta_{n-\hat{\mu},n'} \,.
\end{equation}
Obviously $\left(\hat{U}_\mu\right)^\dagger=\hat{U}_{-\mu}$.
(Note that in terms of these operators the forward and backward covariant 
derivatives are: $\partial_\mu=\hat{U}_\mu -1$,
$\partial_\mu^*=1-\hat{U}_\mu^\dagger$, and
$\partial_\mu^* \partial_\mu=\hat{U}_\mu + \hat{U}_\mu^\dagger -2$.)
The Wilson-Dirac operator at bare mass equal to zero reads:
\begin{equation}
\label{DWdef}
D_{\rm W}=\frac{1}{2} \sum_\mu \gamma_\mu 
\left( \hat{U}_\mu - \hat{U}_\mu^\dagger \right)+ 
\frac{1}{2}r \sum_\mu 
\left( 2 - \hat{U}_\mu - \hat{U}_\mu^\dagger \right) \,.
\end{equation}
It is also useful to introduce the operator $\hat{U}(l)$
of the parallel transport along some path $l=[l_1,l_2,\ldots,l_k]$ 
where $l_i=\pm 1,\ldots,\pm 4$, by
\begin{equation}
\label{Ul}
\hat{U}(l)=\hat{U}_{l_1}\hat{U}_{l_2}\ldots\hat{U}_{l_k} \,.
\end{equation}
In terms of gauge links this is
\begin{equation}
\label{Ul1}
\left( \hat{U}(l)\right)_{nn'}=
\left( U_{l_1}(n)U_{l_2}(n+\hat{l}_1)\ldots \right)\delta_{n+r_l,n'}\,,
\end{equation}
where $r_l=\hat{l}_1+\ldots+\hat{l}_k$ is the offset corresponding
to the path $l$.
Note that $\hat{U}(l)^\dagger=\hat{U}(\bar{l})$ where
$\bar{l}=[-l_k,\ldots,-l_1]$ is the inverse path.
In particular, one has 
\begin{equation}
\label{Ustaple}
\left( \hat{U}([2,1,-2])\right)_{nn'}=
U_2(n)U_1(n+\hat{2})U_2(n+\hat{1})^\dagger \delta_{n+\hat{1},n'}
\end{equation}
for the corresponding staple.

As another example, the Sheikholeslami-Wohlert (or clover) term
\cite{SW}
introduced to cancel the O($a$) artifacts is given
(up to a constant factor) by
\begin{multline}
\label{clover}
i\sigma_{\mu\nu} \left( 
\hat{U}([\mu,\nu,-\mu,-\nu])+\hat{U}([\nu,-\mu,-\nu,\mu])+ \right. \\
\left. \hat{U}([-\mu,-\nu,\mu,\nu])+\hat{U}([-\nu,\mu,\nu,-\mu])
-{\rm h.c.} \right) \,.
\end{multline}

We consider a general form of the Dirac operator
\begin{equation}
\label{Dgen}
D=\sum_A \Gamma_A \sum_l c^A_l \hat{U}(l) \,.
\end{equation}
The Dirac indices are carried by $\Gamma_A=1,\gamma_\mu,\ldots$,
the coordinate and colour indices by the operators $\hat{U}(l)$.
The Dirac operator is determined by the set of paths $l$
the sum runs over, and the coefficients $c^A_l$.

In the case of $D_{\rm W}$, for $\Gamma=1$
one has $l=[1],[-1],\ldots,[-4]$ and $l=[]$ (the empty path 
corresponding to $\hat{U}([])=1$), while for $\Gamma=\gamma_\mu$: 
$l=[\mu]$ and $[-\mu]$. 
In the clover term, eq.~(\ref{clover}) for $\Gamma=i\sigma_{\mu\nu}$:
$l=[\mu,\nu,-\mu,-\nu],\ldots$
(altogether $6\times 4=24$ plaquette products).
As these well known examples indicate, the coefficients $c_l^A$ for 
related paths differ only in relative signs, which are fixed 
by symmetry requirements.

Our aim is to give for all $\Gamma$'s and offsets $r$ on the hypercube
a set of paths and to determine the relative sign
for paths related to each other by symmetry transformations.
We give the general rules for arbitrary offsets and paths as well.

Eqs.~(\ref{pleasant},\ref{ss2}) imply that the coefficients $c_l^A$ in 
eq.~(\ref{Dgen}) are real.
Further, from hermiticity in this language it follows that 
the path $l$ and the opposite path $\bar{l}$ 
(or equivalently,  $\hat{U}(l)$ and  $\hat{U}(l)^\dagger$) 
should enter in the combination
\begin{equation}
\label{GUUd}
\Gamma \left( \hat{U}(l)+\epsilon_\Gamma \hat{U}(l)^\dagger \right) \,,
\end{equation}
where the sign $\epsilon_\Gamma$ is defined by
$\gamma^5 \Gamma^\dagger \gamma^5 =\epsilon_\Gamma \Gamma$,
eq.~(\ref{epsi}).

The symmetry transformations formulated in terms of matrix elements 
in the previous section can be translated to the formalism used here.
The reflections and permutations act on operators $\hat{U}(l)$ in a 
straightforward way. Under a reflection of the axis $\eta$
one has $\hat{U}(l) \to \hat{U}(l')$ where 
$l'_i=-l_i$ if $|l_i|=\eta$ and unchanged otherwise.
Under a permutation $(p_1,p_2,p_3,p_4)$ a component
with $l_i=\pm\mu$ is replaced by $l_i=\pm p_\mu$, as expected.
The number of combined symmetry transformations is $16\times 24=384$.
We denote the action of a transformation $\alpha=1,\ldots,384$
by $\Gamma \to \Gamma^{(\alpha)}$,  
$\hat{U}(l)\to \hat{U}(l^{(\alpha)})$.
Acting on the expression in eq.~(\ref{GUUd}) by all 384 elements
of the symmetry group and adding the resulting operators together,
the sum will satisfy the required symmetry conditions for a Dirac
operator.

Let us introduce the notation
\begin{equation}
\label{sym}
\hat{d}(\Gamma,l)=\frac{1}{\cal N} \sum_\alpha \Gamma^{(\alpha)}
\left( \hat{U}(l^{(\alpha)}) +
\epsilon_\Gamma \hat{U}(l^{(\alpha)})^\dagger \right) \,.
\end{equation}
A general Dirac operator will be a linear combination of such terms.
The normalization factor ${\cal N}$ will be defined below.
The total number of terms in eq.~(\ref{sym}) is $2\times 384=768$. 
Typically, however, the number of different terms which survive 
after the summation is much smaller.
It can happen that for a choice of starting $\Gamma$ and $l$ 
the sum in eq.~(\ref{sym}) is zero. In this case the given 
path does not contribute to the Dirac structure $\Gamma$.

To fix the convention for the overall sign we single out a definite 
term in the sum of eq.~(\ref{sym}) and take its sign to be $+1$.
Denote by $\Gamma_0$, $l_0$ the corresponding quantities of this
reference term, and by $r_0=r(l_0)$ the offset of $l_0$.
This term is specified by narrowing down the set 
$\{ \Gamma^{(\alpha)},l^{(\alpha)} \}$ to a single member
as follows:
\begin{itemize}
\item[a)]
Given an offset $r=r(l)=(r_1,r_2,r_3,r_4)$ the reflections
and permutations create all offsets 
$(\pm r_{p_1},\ \pm r_{p_2},\ \pm r_{p_3},\ \pm r_{p_4})$
where $(p_1,p_2,p_3,p_4)$ is an arbitrary permutation.
We choose for the reference offset $r_0$ the one from the set
$\{ r(l^{(\alpha)}) \}$ which satisfies the relations 
$r_{01} \ge r_{02} \ge r_{03} \ge r_{04} \ge 0$.
\item[b)]
If several $\Gamma$ matrices are generated to {\em this} offset
then choose as $\Gamma_0$ the one which comes first in the natural order,
eq.~(\ref{cord}).
\item[c)]
Consider all the paths $\{ l^{(\alpha)} \}$ having offset $r_0$
and associated with $\Gamma_0$, i.e. $r(l^{(\alpha)})=r_0$
and $\Gamma^{(\alpha)}=\Gamma_0$.
To single out one path $l_0$ from this set, we associate
to a path $l=[l_1,l_2,\ldots,l_k]$ a decimal code $d_1d_2\ldots d_k$
with digits $d_i=l_i$ if $l_i>0$ and $d_i=9+l_i$ for $l_i<0$.
The path with the smallest code will be the reference path $l_0$.
(In other words we take the first in lexical order defined by the
ordering $1,2,3,4,-4,-3,-2,-1$.)
\end{itemize}

Of course, one can take $\Gamma_0$, $l_0$ as the starting $\Gamma$
and $l$, and we shall refer to the expression in eq.~(\ref{sym}) as
$\hat{d}(\Gamma_0,l_0)$ to indicate that it is associated 
to a class rather than to a specific $(\Gamma,l)$.

We turn now to the normalization of $\hat{d}(\Gamma_0,l_0)$.
In general, there will be $K$ {\em different} paths in the set
$\{ l^{(\alpha)} \, | \, \Gamma^{(\alpha)}=\Gamma_0, r(l^{(\alpha)})=r_0\}$,
i.e. corresponding to the same offset $r_0$ and Dirac structure 
$\Gamma_0$.
The normalization is fixed by requiring that the coefficient
of the reference term $\Gamma_0 \hat{U}(l_0)$ is $+1/K$.

Consider a simple example explicitly. Let $r_0=(1,0,0,0)$,
$\Gamma_0=i\sigma_{12}$ and $l_0=[2,1,-2]$. 
The starting term, eq.~(\ref{GUUd}) is 
$i\sigma_{12} ( \hat{U}([2,1,-2])-\hat{U}([2,-1,-2]))$. 
Applying all the 16 different reflections gives 
\begin{equation} 
8 i\sigma_{12}\left( \hat{U}([2,1,-2])-\hat{U}([2,-1,-2]) 
 - \hat{U}([-2,1,2])+\hat{U}([-2,-1,2]) \right) \,.
\end{equation} 
Applying all the permutations on this expression results in: 
\begin{multline}
\label{pelda}
\hat{d}(\Gamma_0,l_0) =  \frac{1}{{\cal N}} \left  \{
16 i\sigma_{12} \left(
\hat{U}([2,1,-2]) - \hat{U}([-2,1,2])  \right)+ \right. \\
   16 i\sigma_{13} \left(
\hat{U}([3,1,-3]) - \hat{U}([-3,1,3])  \right)+ \\
 \left. 16 i\sigma_{14}\left( \hat{U}([4,1,-4]) - \hat{U}([-4,1,4])
 \right) +  \ldots \right\}\,,
\end{multline}
where only the terms with the offset $r=r_0$ are written out explicitly.
Their total number is 96.
The whole generated set has 8 different offsets giving altogether 768 terms. 
Notice the form of the contribution in eq.~(\ref{pelda}). 
There is a common factor (16 in this case) multiplying all the different 
operators. 
Only the tensor elements of the Clifford algebra enter, since we
started with a tensor element. Beyond the common factor the path 
products have a coefficient $\pm 1$. These features are general.
The number of different paths with $\Gamma_0=i\sigma_{12}$
and $r_0=(1,0,0,0)$ is K=2: the paths $[2,1,-2]$ and $[-2,1,2]$).
The normalization factor in this case is ${\cal N}=16\times 2=32$,
so that one has 
$\hat{d}(\Gamma_0,l_0)=\frac{1}{2} i\sigma_{12} \left(
\hat{U}([2,1,-2]) - \hat{U}([-2,1,2]) \right) + \ldots$.

A general Dirac operator is constructed as
\begin{equation}
\label{Dfd}
D= \sum_{\Gamma_0,l_0} f(\Gamma_0,l_0) \hat{d}(\Gamma_0,l_0) \,.
\end{equation}
The coefficients $f(\Gamma_0,l_0)$ are real constants 
or gauge invariant functions of the gauge fields, respecting 
locality, and invariance under the symmetry transformations.
These are the free, adjustable parameters of the Dirac operator.

\section{Tables for offsets on the hypercube}

Choosing offsets and paths to be included in the Dirac operator
is a matter of intuition. It is also influenced by considerations
on CPU time and memory requirements. 
In Tables~\ref{refpath0}-\ref{refpath4}
we give examples of reference paths $l_0$ for offsets on the hypercube
and general Dirac structure.
The first 3 columns give $\Gamma_0$, $l_0$ and the number of paths $K$ 
as defined in the previous section.
The 4th column gives those Clifford basis elements which are
generated in eq.~(\ref{sym}) to the offset $r_0$.

\begin{table}[htbp]
  \begin{center}
    \begin{tabular}{|c|c|c|c|c|}
\hline
$\Gamma_0  $    & ref. path $l_0$             &  $K$ & $\Gamma$'s
generated & $c_X$\\
\hline
$1$           & $[]$                    &   1 & 1 &     $1$, $0$ \\
              & $[1,2,-1,-2]$           &  48 &   &     $1$, $0$ \\
\hline
$\gamma_1$    & $[1,2,-1,-2]$            & 24 & $\gamma^1,\dots,\gamma^4$ & 0\\
\hline
$i\sigma_{12}$ & $[1,2,-1,-2]$     &   8 & $i\sigma_{12},\dots,i\sigma_{34}$ &
$-1$ \\
\hline
$\gamma_5$    & $[1,2,-1,-2,3,4,-3,-4]$ & 384 & $\gamma^5$  & $-1/6$ \\
              & $[1,2,3,4,-1,-2,-3,-4]$ & 384 &             & $-1/6$ \\ 
\hline
$\gamma_1 \gamma_5$ & $[1,2,-1,-2,3,4,-3,-4]$ & 192  & 
                         $\gamma_1\gamma_5,\dots,\gamma_4\gamma_5$ & 0 \\
                    & $[2,1,-2,-1,3,4,-3,-4]$ & 192  & & 0 \\ 
                    & $[2,3,4,-3,-2,-4]$      &  96  & & 0 \\ 
\hline

    \end{tabular}
    \caption{{}Reference paths for different $\Gamma_0$'s for offset (0000).}
    \label{refpath0}
  \end{center}
\end{table}

\begin{table}[htbp]
  \begin{center}
    \begin{tabular}{|c|c|c|c|c|}
\hline
$\Gamma_0$    & ref. path $l_0$             &  $K$ & 
$\Gamma$'s generated &  $c_X$\\
\hline
$1$           & $[1]$                   &   1 & 1   & $8$, $1$\\
              & $[2,1,-2]$              &   6 & & $8$, $1$ \\
\hline
$\gamma_1$    & $[1]$                   &   1 & $\gamma_1$  & $2$ \\
              & $[2,1,-2]$              &   6 & & $2$ \\
\hline
$\gamma_2$    & $[1,2,3,-2,-3]$ & 16 & $\gamma_2,\gamma_3,\gamma_4$ & $0$ \\
\hline
$i\sigma_{12}$& $[2,1,-2]$ & 2 & $i\sigma_{12},i\sigma_{13},i\sigma_{14}$
                                                                    & $4$ \\
\hline
$i\sigma_{23}$ & $[1,2,3,-2,-3]$ & 16 &
                          $i\sigma_{23},i\sigma_{24},i\sigma_{34}$ & $-4$ \\
\hline
$\gamma_5$     & $[2,1,-2,3,4,-3,-4]$ & 96 & $\gamma_5$ & $4/3$ \\
\hline
$\gamma_1 \gamma_5$ & $[2,1,-2,3,4,-3,-4]$  &96   & $\gamma_1\gamma_5$ & $0$\\
\hline
$\gamma_2 \gamma_5$ & $[1,3,4,-3,-4]$  & 16 & 
              $\gamma_2\gamma_5, \gamma_3\gamma_5, \gamma_4 \gamma_5$ & $-2$ \\
\hline
    \end{tabular}
    \caption{{}Reference paths for different $\Gamma_0$'s for offset (1000).}
    \label{refpath1}
  \end{center}
\end{table}

\begin{table}[htbp]
  \begin{center}
    \begin{tabular}{|c|c|c|c|c|}
\hline
$\Gamma_0$    & ref. path $l_0$              &  $K$ & 
$\Gamma$'s generated &  $c_X$ \\
\hline
$1$           & $[1,2]$                   &  2  & 1 & $24$, $6$ \\
\hline
$\gamma_1$    & $[1,2]$     & 2 & $\gamma_1, \gamma_2$ & $12$ \\
\hline
$\gamma_3$    & $[1,3,2,-3]$  & 8 & $\gamma_3, \gamma_4$& $0$ \\
\hline
$i\sigma_{12}$ & $[1,2]$       & 2 & $i\sigma_{12}$ & $-2$ \\
\hline
$i\sigma_{13}$ & $[1,3,2,-3]$ & 4& $i\sigma_{13}, i\sigma_{14},
                              i\sigma_{23}, i\sigma_{24}$ & $0$ \\
\hline
$i\sigma_{34}$ & $[1,2,3,4,-3,-4]$ & 32 & $i\sigma_{34}$ & $-4$ \\
\hline
$\gamma_5$    & $[1,2,3,4,-3,-4]$ & 32 & $\gamma_5$ & $-2$ \\
\hline
$\gamma_1 \gamma_5$ & $[1,2,3,4,-3,-4]$ & 16 & 
                           $\gamma_1\gamma_5, \gamma_2\gamma_5$ & $4$ \\
\hline
$\gamma_3 \gamma_5$ & $[1,4,2,-4]$  &8  & 
                              $\gamma_3\gamma_5, \gamma_4\gamma_5$ & $-4$ \\
\hline
    \end{tabular}
    \caption{{}Reference paths for different $\Gamma_0$'s for offset (1100).}
    \label{refpath2}
  \end{center}
\end{table}

\begin{table}[htbp]
  \begin{center}
    \begin{tabular}{|c|c|c|c|c|}
\hline
$\Gamma_0$    & ref. path $l_0$              &  $K$ &
 $\Gamma$'s generated &  $c_X$ \\
\hline
$1$           & $ [1,2,3]$                   &  6 & 1 & $32$, $12$\\
\hline
$\gamma_1$    & $ [1,2,3]$ &  4 & $\gamma_1, \gamma_2,\gamma_3$ & $24$ \\
\hline
$\gamma_4$    & $[1,2,4,3,-4]$       &  24& $\gamma_4$ & $0$ \\
\hline
$i\sigma_{12}$ & $[1,2,3]$ & 4 & 
       $i\sigma_{12}, i\sigma_{13}, i\sigma_{23}$ & $-8$ \\
\hline
$i\sigma_{14}$ & $[1,4,2,-4,3]$ & 8 & 
       $i\sigma_{14}, i\sigma_{24}, i\sigma_{34}$ & $0$ \\
\hline
$\gamma_5$    & $[1,4,2,-4,3]$ & 12& $\gamma_5$ & $-8/3$ \\
\hline
$\gamma_1 \gamma_5$ & $[1,4,2,-4,3]$  & 8 & $\gamma_1 \gamma_5, 
\gamma_2 \gamma_5, \gamma_3 \gamma_5$ & $8$\\
\hline
$\gamma_4 \gamma_5$ & $[1,2,3]$  & 6  & $\gamma_4 \gamma_5$ & $-4/3$ \\
\hline
    \end{tabular}
    \caption{{}Reference paths for different $\Gamma_0$'s for offset (1110).}
    \label{refpath3}
  \end{center}
\end{table}

\begin{table}[htbp]
  \begin{center}
    \begin{tabular}{|c|c|c|c|c|}
\hline
$\Gamma_0  $    & ref. path $l_0$             &  $K$ & $\Gamma$'s
generated & $c_X$\\
\hline
$1$           & $[1,2,3,4]$                    &   24 & 1 & $16$, $8$  \\
\hline
$\gamma_1$    & $[1,2,3,4]$ &  12 & $\gamma^1,\dots, \gamma^4$ & $16$ \\
\hline
$i\sigma_{12}$ & $[1,2,3,4]$     &  8 & 
     $i\sigma_{12},\dots,i\sigma_{34}$ & $-8$ \\
\hline
$\gamma_5$    & $[1,2,3,4]$ & 24 & $\gamma^5$  & $-2/3$ \\
\hline
$\gamma_1 \gamma_5$ & $[1,2,3,4]$  &  12 & $\gamma_1 \gamma_5,\dots,
\gamma_4 \gamma_5$    & $8/3$ \\
\hline

    \end{tabular}
    \caption{{}Reference paths for different $\Gamma_0$'s for offset (1111).}
    \label{refpath4}
  \end{center}
\end{table}

It is also of interest how the given Dirac operator behaves
for smooth gauge fields, i.e. to obtain the leading terms in the naive
continuum limit.
The expression in eq.~(\ref{sym})
generated by given $\Gamma_0$ and $l_0$ contributes in this limit
to one of the expressions (of type S,V,T,P,A) below
\begin{multline}
  \label{clim}
  \left( \bar{c}_{\rm S} + c_{\rm S} \partial^2 \right) \,; \quad
  c_{\rm V} \gamma_\mu \partial_\mu \,; \quad
  c_{\rm T} \frac{1}{2} \sigma_{\mu\nu} F_{\mu\nu}\,; \\
  c_{\rm P} \frac{1}{4} \gamma_5 
       \epsilon_{\mu\nu\rho\sigma}F_{\mu\nu}F_{\rho\sigma}\,; \quad
  c_{\rm A} i \gamma_\mu \gamma_5 
     \epsilon_{\mu\nu\rho\sigma}\partial_\nu F_{\rho\sigma} \,.
\end{multline}
Here $\partial_\mu$ is the covariant derivative in the continuum.
The coefficients $c_X(\Gamma_0,l_0)$, which determine the continuum 
behaviour of the Dirac operator, are presented in the last column 
of Tables~\ref{refpath0}-\ref{refpath4}.
For $\Gamma_0=1$ the first and second entries correspond to 
$\bar{c}_{\rm S}$ and $c_{\rm S}$, respectively.
We list their meaning below.

Introduce the notation
\begin{equation}
C_X=\sum_{\Gamma_0,l_0} f(\Gamma_0,l_0) c_X(\Gamma_0,l_0) \,.
\end{equation}

The bare mass $m_0$ is given by
\begin{equation}
  \label{eq:M}
  \bar{C}_{\rm S}=m_0 \,.
\end{equation}

The normalization condition on the $D\sim \gamma_\mu \partial_\mu$
term in $D$ gives
\begin{equation}
  \label{eq:V}
 C_{\rm V}=1 \,.
\end{equation}

The O($a$) tree level Symanzik condition reads
\begin{equation}
  \label{eq:ST}
C_{\rm S} +  C_{\rm T}=0 \,.
\end{equation}

The coefficient $c_{\rm P}$ is interesting if the parametrization
attempts to describe (approximately) a GW fermion.
In this case it is related to the topological charge density
$q(n)={\rm Tr}(\gamma_5 D R)_{nn}$,
\begin{equation}
  \label{eq:P}
  \sum_{\Gamma_0,l_0} f(\Gamma_0,l_0) c_{\rm P}(\Gamma_0,l_0) 
R^{\rm free}(r_0)=
\frac{1}{32\pi^2} \,.
\end{equation}
(If $f(\Gamma_0,l_0)$ is not a constant, then its continuum
limit is understood here.)
Of course, in the expressions above $\Gamma_0$ should be of the
corresponding type (S,T,\ldots).
The quantity $R^{\rm free}(r_0)$ is given by $R(n,n+r_0;U=1)$ 
of the GW relation eq.(\ref{gw}).

\section{The list of factorized contributions}

As the tables~\ref{refpath0}-\ref{refpath4} show the number of paths, in 
particularly for the offset $r_0=(0,0,0,0)$, is large. It is important to
calculate the path products efficiently. An obvious method is to factorize the
paths, i.e. to write the sum of a large number of paths as a product of sums
over shorter paths. We shall try to factorize in such a way that these shorter
paths are mainly plaquette, or staple products.

Define the plaquette products (i.e. the operators of parallel
transport along a plaquette) in the following way:
\begin{equation}
\label{Plaq}
P_{l_1 ,l_2}=\hat{U}([l_1,l_2,-l_1,-l_2]) \,,
\end{equation}
where $l_i=\pm 1,\ldots,\pm4$.
Their hermitian conjugate is given by $(P_{l_1 ,l_2})^\dagger=P_{l_2 ,l_1}$. 
Reflections and permutations act on them in an obvious way.

We also define the staple products as
\begin{equation}
\label{Stap}
S_{l_1 ,l_2}=\hat{U}([l_1,l_2,-l_1]) \,.
\end{equation}
We have $(S_{l_1 ,l_2})^\dagger = S_{l_1 ,-l_2}$.

To describe the shortest path to an offset we introduce the notation
\begin{eqnarray}
\label{line}
& & V_{l_1}=\hat{U}([l_1]) \,, \nonumber \\
& & V_{l_1 ,l_2}=\hat{U}([l_1,l_2]) \,, \nonumber \\
& & V_{l_1 ,l_2,l_3}=\hat{U}([l_1,l_2,l_3]) \,,  \\
& & V_{l_1 ,l_2,l_3,l_4}=\hat{U}([l_1,l_2,l_3,l_4]) \,. \nonumber 
\end{eqnarray}

Introduce the following linear combinations
transforming in a simple way under reflections:
\begin{eqnarray}
\label{ppp}
& &P^{(++)}_{\mu\nu} = 
       P_{\mu,\nu}+P_{\mu,-\nu}+P_{-\mu,\nu}+P_{-\mu,-\nu} \,, \nonumber \\
& &P^{(+-)}_{\mu\nu} = 
       P_{\mu,\nu}-P_{\mu,-\nu}+P_{-\mu,\nu}-P_{-\mu,-\nu} \,,  \\
& &P^{(-+)}_{\mu\nu} = 
       P_{\mu,\nu}+P_{\mu,-\nu}-P_{-\mu,\nu}-P_{-\mu,-\nu} \,, \nonumber \\
& &P^{(--)}_{\mu\nu} = 
       P_{\mu,\nu}-P_{\mu,-\nu}-P_{-\mu,\nu}+P_{-\mu,-\nu} \,. \nonumber 
\end{eqnarray}
The signs in the superscript denote the parity for reflections
of the $\mu$, $\nu$ axes, respectively.
Hermitian conjugation acts as interchanging both upper and lower
indices, permutations simply by $\mu \to p_\mu$, $\nu \to p_\nu$.

It is also useful to denote combinations which are 
symmetric/antisymmetric with respect to interchanging the axes:
\begin{eqnarray}
\label{psa}
& &P^{\rm (sym)}_{\mu\nu} = 
       P^{(++)}_{\mu,\nu}+P^{(++)}_{\nu,\mu}
= P^{(++)}_{\mu,\nu}+{\rm h.c.} \,, \\
& &P^{\rm (as)}_{\mu\nu} = 
       P^{(--)}_{\mu,\nu}-P^{(--)}_{\nu,\mu}
= P^{(--)}_{\mu,\nu}-{\rm h.c.} \,. \nonumber 
\end{eqnarray}

For the staples we write
\begin{eqnarray}
\label{sd1}
& &S^{(\nu +)}_{\mu}  = S_{\nu,\mu}  + S_{-\nu,\mu} \,, \nonumber \\
& &S^{(\nu -)}_{\mu}  = S_{\nu,\mu}  - S_{-\nu,\mu} \,, \\
& &S^{(\nu +)}_{-\mu} = S_{\nu,-\mu} + S_{-\nu,-\mu}\,, \nonumber \\
& &S^{(\nu -)}_{-\mu} = S_{\nu,-\mu} - S_{-\nu,-\mu}\,. \nonumber 
\end{eqnarray}
The subscript $\pm\mu$ denotes the direction of the staple, 
the superscript specifies the plane $\mu\nu$ and the parity
in $\nu$.

For a line of length 2 we have
\begin{eqnarray}
\label{line1}
& &V^{(++)}_{\mu\nu} = 
       V_{\mu,\nu}+V_{\mu,-\nu}+V_{-\mu,\nu}+V_{-\mu,-\nu} \,, \nonumber \\
& &V^{(+-)}_{\mu\nu} = 
       V_{\mu,\nu}-V_{\mu,-\nu}+V_{-\mu,\nu}-V_{-\mu,-\nu} \,,  \\
& &V^{(-+)}_{\mu\nu} = 
       V_{\mu,\nu}+V_{\mu,-\nu}-V_{-\mu,\nu}-V_{-\mu,-\nu} \,, \nonumber \\
& &V^{(--)}_{\mu\nu} = 
       V_{\mu,\nu}-V_{\mu,-\nu}-V_{-\mu,\nu}+V_{-\mu,-\nu} \,, \nonumber 
\end{eqnarray}
and analogously for longer lines. In analogy to eq.~(\ref{psa})
we introduce the completely (anti)symmetric combinations 
$V^{\rm (sym)}_{\mu\nu\ldots}$, $V^{\rm (as)}_{\mu\nu\ldots}$
for the line products.

Define also the bent rectangle,
\begin{equation}
B_{l_1,l_2,l_3}=\hat{U}([l_1,l_2,l_3,-l_2,-l_1,-l_3]) \,,
\end{equation}
and $B^{(---)}_{\nu\rho\sigma}$ which is odd in all its arguments
(as a sum of 8 terms, defined analogously to $P^{(--)}_{\mu\nu}$).
Similarly, we introduce the ``4d plaquette'' 
\begin{equation}
Q_{l_1,l_2,l_3,l_4}=\hat{U}([l_1,l_2,l_3,l_4,-l_1,-l_2,-l_3,-l_4]) \,,
\end{equation}
and the odd combination $Q_{l_1,l_2,l_3,l_4}^{(----)}$.

To simplify the expressions below, we require that all the directions 
entering $P_{l_1,l_2}$, $S_{l_1,l_2}$, $V_{l_1,l_2,\ldots}$, etc.
are different, i.e. they are taken to be zero if e.g. $|l_1|=|l_2|$.

Below we list $\hat{d}(\Gamma_0,l_0)$
for different choices of $\Gamma_0$ and $l_0$.
The notation $\left. \sum \right.'$ indicates that  all indices 
in the corresponding sum are taken to be different.

\subsection{The offset $r_0=(0,0,0,0)$}

\noindent{$\Gamma_0=1$, $l_0=[\,]$}
\begin{equation}
\label{g0a_0}
1 \,.
\end{equation}

\noindent{$\Gamma_0=1$, $l_0=[1,2,-1,-2]$}
\begin{equation}
\label{g0b_0}
\frac{1}{48}  \sum_{\mu<\nu} P^{\rm (sym)}_{\mu\nu} \,.
\end{equation}

\noindent{$\Gamma_0 = \gamma_1$, $l_0=[1,2,-1,-2]$}
\begin{equation}
\label{g1_0}
\frac{1}{24} \left. \sum_{\mu\nu} \right.'
\gamma_\mu \left( P^{(-+)}_{\mu\nu} - \mbox{h.c.} \right) \,.
\end{equation}

\noindent{$\Gamma_0= i\sigma_{12}$, $l_0=[1,2,-1,-2]$}
\begin{equation}
\label{s12_0}
\frac{1}{8} \sum_{\mu<\nu} i\sigma_{\mu\nu} P^{\rm (as)}_{\mu\nu} \,.
\end{equation}

\noindent{$\Gamma_0= \gamma_5$, $l_0=[1,2,-1,-2,3,4,-3,-4]$}
\begin{equation}
\label{g5a_0}
 \frac{1}{384} \gamma_5 \left. \sum_{\mu\nu\rho\sigma} \right.'
\frac{1}{4} \epsilon_{\mu\nu\rho\sigma}
 P^{\rm (as)}_{\mu\nu}P^{\rm (as)}_{\rho\sigma} \,.
\end{equation}

\noindent{$\Gamma_0= \gamma_5$, $l_0=[1,2,3,4,-1,-2,-3,-4]$}
\begin{equation}
\label{g5b_0}
\frac{1}{384}  \gamma_5 \left. \sum_{\mu\nu\rho\sigma} \right.'
\epsilon_{\mu\nu\rho\sigma} Q_{\mu\nu\rho\sigma}^{(----)} \,.
\end{equation}

\noindent{$\Gamma_0= \gamma_1\gamma_5$, $l_0=[1,2,-1,-2,3,4,-3,-4]$}
\begin{equation}
\label{g15a_0}
\frac{1}{192} \left. \sum_{\mu\nu\rho\sigma} \right.'
\gamma_\mu\gamma_5 \frac{1}{2}\epsilon_{\mu\nu\rho\sigma}
\left( P^{(+-)}_{\mu\nu}P^{\rm (as)}_{\rho\sigma} + \mbox{h.c.} \right) \,.
\end{equation}

\noindent{$\Gamma_0= \gamma_1\gamma_5$, $l_0=[2,1,-2,-1,3,4,-3,-4]$}
\begin{equation}
\label{g15b_0}
 \frac{1}{192} \left. \sum_{\mu\nu\rho\sigma} \right.'
\gamma_\mu\gamma_5 \frac{1}{2} \epsilon_{\mu\nu\rho\sigma}
\left( P^{(-+)}_{\nu\mu}P^{\rm (as)}_{\rho\sigma} + \mbox{h.c.} \right) \,.
\end{equation}

\noindent{$\Gamma_0= \gamma_1\gamma_5$, $l_0=[2,3,4,-3,-2,-4]$}
\begin{equation}
\label{g15c_0}
  \frac{1}{96} \left. \sum_{\mu\nu\rho\sigma} \right.'
\gamma_\mu\gamma_5 \epsilon_{\mu\nu\rho\sigma}
\left( B^{(---)}_{\nu\rho\sigma}  + \mbox{h.c.} \right) \,.
\end{equation}

\subsection{The offset $r_0=(1,0,0,0)$}

\noindent{$\Gamma_0 = 1$,  $l_0=[1]$}\\
\begin{equation}
\label{g0a_1}
\sum_\mu \left( V_\mu + V_{-\mu} \right) \,.
\end{equation}

\noindent{$\Gamma_0 = 1$, $l_0=[2,1,-2]$}\\
\begin{equation}
\label{g0b_1}
\frac{1}{6} \left. \sum_{\mu\nu} \right.'
 \left( S^{(\nu,+)}_\mu + S^{(\nu,+)}_{-\mu} \right) \,.
\end{equation}

\noindent{$\Gamma_0 = \gamma_1$, $l_0=[1]$}\\
\begin{equation}
\label{g1a_1}
\sum_\mu \gamma_\mu \left( V_\mu - V_{-\mu} \right) \,.
\end{equation}

\noindent{$\Gamma_0 = \gamma_1$, $l_0=[2,1,-2]$}\\
\begin{equation}
\label{g1b_1}
\frac{1}{6} \left. \sum_{\mu\nu} \right.' \gamma_\mu 
\left( S^{(\nu,+)}_\mu - S^{(\nu,+)}_{-\mu} \right) \,.
\end{equation}

\noindent{$\Gamma_0 = \gamma_2$, $l_0=[1,2,3,-2,-3]$}\\
\begin{equation}
\label{g2_1}
\frac{1}{16} 
\left. \sum_{\mu\nu\rho}\right.' \, \gamma_\nu
\left(  V_\mu P^{(-+)}_{\nu\rho} - P^{(+-)}_{\rho\nu} V_\mu - {\rm h.c.}
\right) \,.
\end{equation}

\noindent{$\Gamma_0 = i\sigma_{12}$, $l_0=[2,1,-2]$}\\
\begin{equation}
\label{s12_1}
\frac{1}{2} \left. \sum_{\mu\nu} \right.'  i\sigma_{\mu\nu} \frac{1}{2}
\left(  S^{(\nu,-)}_\mu - S^{(\mu,-)}_\nu - \mbox{h.c.} \right) \,.
\end{equation}

\noindent{$\Gamma_0 = i\sigma_{23}$, $l_0=[1,2,3,-2,-3]$}\\
\begin{equation}
\label{s23a_1}
\frac{1}{16} \left. \sum_{\mu\nu\rho} \right.' 
i\sigma_{\mu\nu} \frac{1}{2} 
 \left(  V_\rho P^{\rm (as)}_{\mu\nu} + 
 P^{\rm (as)}_{\mu\nu} V_\rho-\mbox{h.c.} \right) \,.
\end{equation}


\noindent{$\Gamma_0 = \gamma_5$, $l_0=[2,1,-2,3,4,-3,-4]$}\\
\begin{equation}
\label{g5_1}
\frac{1}{96} \gamma_5 \left. \sum_{\mu\nu\rho\sigma} \right.'
\frac{1}{2} \epsilon_{\mu\nu\rho\sigma}
\left( S^{(\nu,-)}_\mu P^{\rm (as)}_{\rho\sigma}
 + P^{\rm (as)}_{\rho\sigma} S^{(\nu,-)}_\mu + \mbox{h.c.} \right) \,.
\end{equation}

\noindent{$\Gamma_0 = \gamma_1\gamma_5$, $l_0=[2,1,-2,3,4,-3,-4]$}\\
\begin{equation}
\label{g15_1}
\frac{1}{96} \left. \sum_{\mu\nu\rho\sigma} \right.' 
 \gamma_\mu\gamma_5 \frac{1}{2} \epsilon_{\mu\nu\rho\sigma}
\left( S^{(\nu,-)}_\mu P^{\rm (as)}_{\rho\sigma}
 - P^{\rm (as)}_{\rho\sigma} S^{(\nu,-)}_\mu + \mbox{h.c.} \right) \,.
\end{equation}

\noindent{$\Gamma_0 = \gamma_2 \gamma_5$, $l_0=[1,3,4,-3,-4]$}\\
\begin{equation}
\label{g25a_1}
\frac{1}{16} \left. \sum_{\mu\nu\rho\sigma} \right.' 
\gamma_\nu\gamma_5 \frac{1}{2} 
\epsilon_{\mu\nu\rho\sigma}
\left( V_\mu P^{\rm (as)}_{\rho\sigma}
 + P^{\rm (as)}_{\rho\sigma} V_\mu + \mbox{h.c.} \right) \,.
\end{equation}


\subsection{The offset $r_0=(1,1,0,0)$}

\noindent{$\Gamma_0=1$, $l_0=[1,2]$}\\
\begin{equation}
\label{g0a_2}
\frac{1}{2} \sum_{\mu<\nu} V_{\mu\nu}^{\rm (sym)} \,.
\end{equation}


\noindent{$\Gamma_0=\gamma_1$, $l_0=[1,2]$}\\
\begin{equation}
\label{g1a_2}
\frac{1}{2} \left. \sum_{\mu\nu} \right.' 
\gamma_\mu \left( V_{\mu\nu}^{(-+)} - \mbox{h.c.} \right) \,.
\end{equation}


\noindent{$\Gamma_0=\gamma_3$, $l_0=[1,3,2,-3]$}\\
\begin{equation}
\label{g3_2}
\frac{1}{8} \left. \sum_{\mu\nu\rho} \right.'
\gamma_\rho \left( 
  V_{\mu} S_{\nu}^{(\rho,-)} + V_{-\mu} S_{\nu}^{(\rho,-)}
  - S_{\mu}^{(\rho,-)} V_{\nu} - S_{\mu}^{(\rho,-)} V_{-\nu}
 - \mbox{h.c.} \right) \,.
\end{equation}

\noindent{$\Gamma_0=i\sigma_{12}$, $l_0=[1,2]$}\\
\begin{equation}
\label{s12a_2}
\frac{1}{2} \sum_{\mu < \nu}  i\sigma_{\mu\nu} V_{\mu\nu}^{\rm (as)} \,.
\end{equation}


\noindent{$\Gamma_0=i\sigma_{13}$, $l_0=[1,3,2,-3]$}\\
\begin{equation}
\label{s13_2}
\frac{1}{4} \left. \sum_{\mu\nu\rho} \right.'
i\sigma_{\mu\rho} \left(
  V_{\mu} S_{\nu}^{(\rho,-)} - V_{-\mu} S_{\nu}^{(\rho,-)}
 +V_{\mu} S_{-\nu}^{(\rho,-)} - V_{-\mu} S_{-\nu}^{(\rho,-)}
 - \mbox{h.c.} \right) \,.
\end{equation}

\noindent{$\Gamma_0=i\sigma_{34}$, $l_0=[1,2,3,4,-3,-4]$}\\
\begin{equation}
\label{s34a_2}
\frac{1}{32} \left. \sum_{\mu\nu\rho\sigma} \right.'
i\sigma_{\rho\sigma} \frac{1}{2} \left(
V_{\mu\nu}^{(++)} P_{\rho\sigma}^{\rm (as)} - \mbox{h.c.} \right) \,.
\end{equation}


\noindent{$\Gamma_0=\gamma_5$, $l_0=[1,2,3,4,-3,-4]$}\\
\begin{equation}
\label{g5a_2}
\frac{1}{32} \gamma_5 \left. \sum_{\mu\nu\rho\sigma} \right.'
 \frac{1}{2} \epsilon_{\mu\nu\rho\sigma} \left(
V_{\mu\nu}^{(--)} P_{\rho\sigma}^{\rm (as)} + \mbox{h.c.} \right) \,.
\end{equation}


\noindent{$\Gamma_0=\gamma_1\gamma_5$, $l_0=[1,2,3,4,-3,-4]$}\\
\begin{equation}
\label{g15a_2}
\frac{1}{16} \left. \sum_{\mu\nu\rho\sigma} \right.'
\gamma_\mu \gamma_5 \frac{1}{2} \epsilon_{\mu\nu\rho\sigma} \left(
V_{\mu\nu}^{(+-)} P_{\rho\sigma}^{\rm (as)} + \mbox{h.c.} \right) \,.
\end{equation}


\noindent{$\Gamma_0=\gamma_3\gamma_5$, $l_0=[1,4,2,-4]$}\\
\begin{multline}
\label{g35_2}
\frac{1}{8} \left. \sum_{\mu\nu\rho\sigma} \right.'
\gamma_\rho \gamma_5 \epsilon_{\mu\nu\rho\sigma} \left(
  V_{\mu} S_{\nu}^{(\sigma,-)}  - V_{-\mu} S_{\nu}^{(\sigma,-)} \right. \\
\left. -V_{\mu} S_{-\nu}^{(\sigma,-)} + V_{-\mu} S_{-\nu}^{(\sigma,-)}
 + \mbox{h.c.} \right) \,.
\end{multline}

\subsection{The offset $r_0=(1,1,1,0)$}

\noindent{$\Gamma_0 = 1$,  $l_0=[1,2,3]$}\\
\begin{equation}
\label{g0_3}
\frac{1}{6} \sum_{\mu<\nu<\rho} V_{\mu\nu\rho}^{\rm (sym)} \,.
\end{equation}

\noindent{$\Gamma_0 = \gamma_1$,  $l_0=[1,2,3]$}\\
\begin{equation}
\label{g1_3}
\frac{1}{4} \left. \sum_{\mu\nu\rho} \right.'
\gamma_\mu \left(  V_{\mu\nu\rho}^{(-++)}+V_{\nu\rho\mu}^{(++-)} \right) \,.
\end{equation}

\noindent{$\Gamma_0 = \gamma_4$,  $l_0=[1,2,4,3,-4]$}\\
\begin{equation}
\label{g4_3}
\frac{1}{24} \left. \sum_{\mu\nu\rho\sigma} \right.'
 \gamma_\nu \frac{1}{2}
\left( 
  V^{\rm (sym)}_{\rho\sigma}S_\mu^{(\nu,-)}
+ V^{\rm (sym)}_{\rho\sigma}S_{-\mu}^{(\nu,-)} - \mbox{h.c.}
\right) \,.
\end{equation}

\noindent{$\Gamma_0 = i\sigma_{12}$,  $l_0=[1,2,3]$}\\
\begin{equation}
\label{s12_3}
\frac{1}{4} \left. \sum_{\mu\nu\rho} \right.'
i\sigma_{\mu\nu} \frac{1}{2} \left(
V_{\mu\nu\rho}^{(--+)}+V_{\rho\mu\nu}^{(+--)} -  \mbox{h.c.}
\right) \,.
\end{equation}

\noindent{$\Gamma_0 = i\sigma_{14}$,  $l_0=[1,4,2,-4,3]$}\\
\begin{multline}
\label{s14_3}
\frac{1}{8} \left. \sum_{\mu\nu\rho\sigma} \right.'
i\sigma_{\mu\nu} \left( 
 V_\mu    S_\rho^{(\nu,-)}V_\sigma    + V_\sigma S_\rho^{(\nu,-)}V_\mu
-V_{-\mu} S_\rho^{(\nu,-)}V_\sigma    - V_\sigma S_\rho^{(\nu,-)}V_{-\mu}
                                                                  \right. \\
\left.
+V_\mu S_\rho^{(\nu,-)}V_{-\sigma}    + V_{-\sigma} S_\rho^{(\nu,-)}V_\mu
-V_{-\mu} S_\rho^{(\nu,-)}V_{-\sigma} - V_{-\sigma} S_\rho^{(\nu,-)}V_{-\mu}
- \mbox{h.c.} \right) \,.
\end{multline}

\noindent{$\Gamma_0 = \gamma_5$,  $l_0=[1,4,2,-4,3]$}\\
\begin{multline}
\label{g5_3}
\frac{1}{12} \gamma_5 \left. \sum_{\mu\nu\rho\sigma} \right.'
\frac{1}{2} \epsilon_{\mu\nu\rho\sigma}
\left( 
 V_\mu    S_\rho^{(\nu,-)}V_\sigma    - V_\sigma S_\rho^{(\nu,-)}V_\mu
-V_{-\mu} S_\rho^{(\nu,-)}V_\sigma    + V_\sigma S_\rho^{(\nu,-)}V_{-\mu}
                                                                  \right. \\
\left.
-V_\mu S_\rho^{(\nu,-)}V_{-\sigma}    + V_{-\sigma} S_\rho^{(\nu,-)}V_\mu
+V_{-\mu} S_\rho^{(\nu,-)}V_{-\sigma} - V_{-\sigma} S_\rho^{(\nu,-)}V_{-\mu}
+ \mbox{h.c.} \right) \,.
\end{multline}

\noindent{$\Gamma_0 = \gamma_1\gamma_5$,  $l_0=[1,4,2,-4,3]$}\\
\begin{multline}
\label{g15_3}
\frac{1}{8} \left. \sum_{\mu\nu\rho\sigma} \right.'
\gamma_\mu \gamma_5 \epsilon_{\mu\nu\rho\sigma}
\left( 
 V_\mu    S_\rho^{(\nu,-)}V_\sigma    + V_\sigma S_\rho^{(\nu,-)}V_\mu
+V_{-\mu} S_\rho^{(\nu,-)}V_\sigma    + V_\sigma S_\rho^{(\nu,-)}V_{-\mu}
                                                                  \right. \\
\left.
-V_\mu S_\rho^{(\nu,-)}V_{-\sigma}    - V_{-\sigma} S_\rho^{(\nu,-)}V_\mu
-V_{-\mu} S_\rho^{(\nu,-)}V_{-\sigma} - V_{-\sigma} S_\rho^{(\nu,-)}V_{-\mu}
+ \mbox{h.c.} \right) \,.
\end{multline}

\noindent{$\Gamma_0 = \gamma_4\gamma_5$,  $l_0=[1,2,3]$}\\
\begin{equation}
\label{g45_3}
\frac{1}{6} \left. \sum_{\mu\nu\rho\sigma} \right.'
\gamma_\sigma \gamma_5 \frac{1}{2} \epsilon_{\mu\nu\rho\sigma}
\left( V_{\mu\nu\rho}^{(---)} + \mbox{h.c.} \right) \,.
\end{equation}

\subsection{The offset $r_0=(1,1,1,1)$}

\noindent{$\Gamma_0 = 1$,  $l_0=[1,2,3,4]$}\\
\begin{equation}
\label{g0_4}
\frac{1}{24} V_{1234}^{\rm (sym)} \,.
\end{equation}

\noindent{$\Gamma_0 = \gamma_1$,  $l_0=[1,2,3,4]$}\\
\begin{equation}
\frac{1}{12} \left. \sum_{\mu\nu\rho\sigma} \right.'
\gamma_\mu \left( V_{\mu\nu\rho\sigma}^{(-+++)} - \mbox{h.c.} \right) \,.
\label{g1_4}
\end{equation}

\noindent{$\Gamma_0 = i\sigma_{12}$,  $l_0=[1,2,3,4]$}\\
\begin{equation}
\label{s12_4}
\frac{1}{8} \left. \sum_{\mu\nu\rho\sigma} \right.'
i\sigma_{\mu\nu} \left( V_{\mu\nu\rho\sigma}^{(--++)}- \mbox{h.c.} \right) \,.
\end{equation}

\noindent{$\Gamma_0 = \gamma_5$,  $l_0=[1,2,3,4]$}\\
\begin{equation}
\label{g5_4}
\frac{1}{24} \gamma_5 \left. \sum_{\mu\nu\rho\sigma} \right.'
\frac{1}{2} \epsilon_{\mu\nu\rho\sigma} 
\left( V_{\mu\nu\rho\sigma}^{(----)} + \mbox{h.c.} \right)\,.
\end{equation}

\noindent{$\Gamma_0 = \gamma_1\gamma_5$,  $l_0=[1,2,3,4]$}\\
\begin{equation}
\label{g15_4}
\frac{1}{12} \left. \sum_{\mu\nu\rho\sigma} \right.'
\gamma_\mu \gamma_5  \epsilon_{\mu\nu\rho\sigma} 
\left( V_{\mu\nu\rho\sigma}^{(+---)} + \mbox{h.c.} \right) \,.
\end{equation}

\section{Concluding remarks}
We have given a construction of a generalized Dirac operator
with the required symmetry properties.
For every offset on the hypercube and every element
of the Clifford algebra we have provided at least one example of a 
contribution to the Dirac operator.

Theoretical or computational considerations should determine the
importance of a particular term, which could require the reduction 
or extension of the set contained in Tables 1-5. We plan to 
optimize the parameterization to approximate the fixed-point 
Dirac operator, in order to have good scaling and chiral properties. 

{\bf Acknowledgment}: We thank Philipp R\"ufenacht for useful discussions.

\eject
\end{document}